\documentclass[twocolumn,aps,prb,showpacs,floatfix,epsfig]{revtex4}

\usepackage{graphicx}%
\usepackage{dcolumn}
\usepackage{bm}%

\newcommand \be{\begin{equation}}
\newcommand \ee{\end{equation}}
\newcommand \ba{\begin{eqnarray}}
\newcommand \ea{\end{eqnarray}}
\begin{document}
\title
{\bf Impact ionization fronts in semiconductors: superfast
propagation due to ``nonlocalized'' preionization}
\author{Pavel Rodin$^{1}$ \cite{EMAIL}, Andrey Minarsky$^{2}$ and Igor Grekhov$^{1}$}
\affiliation
{$^1$ Ioffe Physicotechnical Institute, Politechnicheskaya 26,
194021, St.-Petersburg, Russia,\\
$^2$ Physico-Technical High School of Russian Academy of Science,
Khlopina 8-3,194021, St.-Petersburg, Russia}

\setcounter{page}{1}
\date{\today}

\hyphenation{cha-rac-te-ris-tics}
\hyphenation{se-mi-con-duc-tor}
\hyphenation{fluc-tua-tion}
\hyphenation{fi-la-men-ta-tion}
\hyphenation{self--con-sis-tent}
\hyphenation{cor-res-pon-ding}
\hyphenation{con-duc-ti-vi-ti-tes}


\begin{abstract}
We discuss a new mode of ionization front passage in semiconductor structures.
The front of avalanche ionization
propagates into an intrinsic semiconductor with a constant
electric field $E_{\rm m}$ in presence of a small
concentration of free nonequilibrium carriers - so called preionization.
We show that if  the profile of these initial carriers decays in the direction
of the front propagation with a characteristic exponent $\lambda$,
the front velocity is determined by
$v_f \approx 2 \beta_{\rm m}/\lambda$, where $\beta_{\rm m} \equiv \beta(E_{\rm m})$
is the corresponding ionization frequency. By a proper choice of the preionization profile
one can achieve front velocities $v_f$ that exceed the saturated
drift velocity $v_s$ by several orders
of magnitude even in moderate electric fields.
Our propagation mechanism differs from the one for well-known TRAPATT fronts.
Finally, we discuss physical reasons for the appearance of preionization profiles
with slow spatial decay.
\end{abstract}



\maketitle

Propagation of impact ionization fronts in semiconductor
structures represents a spectacular nonlinear effect \cite{LEV05,PRA68,Si,GaAs}
which has important applications in pulse power electronics. \cite{applications}
In reverse-biased $p^+$-$n$-$n^+$ diode structures ionizing fronts propagate faster
than the saturated drift velocity $v_s$. \cite{LEV05,PRA68,Si,GaAs} Such superfast propagation
is possible due to the presence of small concentrations $n_0$,$p_0$ of
free electrons and holes in the depleted region. These
free carriers which initiate an avalanche multiplication are often
coined as ``pre-ionization'' of the medium. \cite{LEV05,DYA}
According to the conventional concept of ionization fronts
in TRAPATT (TRAped Plasma Avalanche Triggered Transit) diodes
\cite{DEL70,ROD07} the avalanche multiplication
occurs within the ionization zone of length $\ell_f=
\varepsilon \varepsilon_0 (E_{\rm m}-E_b)/q N_d$ where electric
field exceeds the effective threshold of impact ionization
$E_b$ (Fig.~1, curve 1). This length is finite due to the slope of the electric field in
the $n$ base $dE/dx=q N_d/ \varepsilon \varepsilon_0$ which depends
on the doping level $N_d$ (note that $n_0,p_0 \ll N_d$). The finiteness of the ionization
zone $\ell_f$ prevents a uniform avalanche multiplication in the whole
$n$ base and thus ensures the existence of the traveling front
mode of avalanche breakdown. However, this concept is not applicable to
$p$-$i$-$n$ structures with intrinsic ($N_d=0$) base (Fig.~1, curve 2)
as well as to short overvoltaged structures because in both cases
$E>E_b$ in the whole $n$ base (Fig.~1, curve 3). On the basis of TRAPATT-like
front concept one would expect that in these two cases pre-ionization
of the high-field region triggers quasiuniform breakdown ruining
the traveling front mode.

In this paper we argue that superfast impact ionization fronts are nevertheless
possible in $p$-$i$-$n$ structures where $E>E_b$ everywhere in the
high-field region providing the concentration profile of initial
carriers $n_0(x),p_0(x)$ decays in the direction of front
propagation. The propagation mechanism of such front is completely
different from the conventional TRAPATT-like front. We find the front
velocity analytically and show that it is controlled by the slope of
pre-ionization profile, and that it can exceed $v_s$ by several orders of magnitude.
\begin{figure}
\begin{center}
\includegraphics[width=5.0 cm,height=6.75 cm,angle=270]{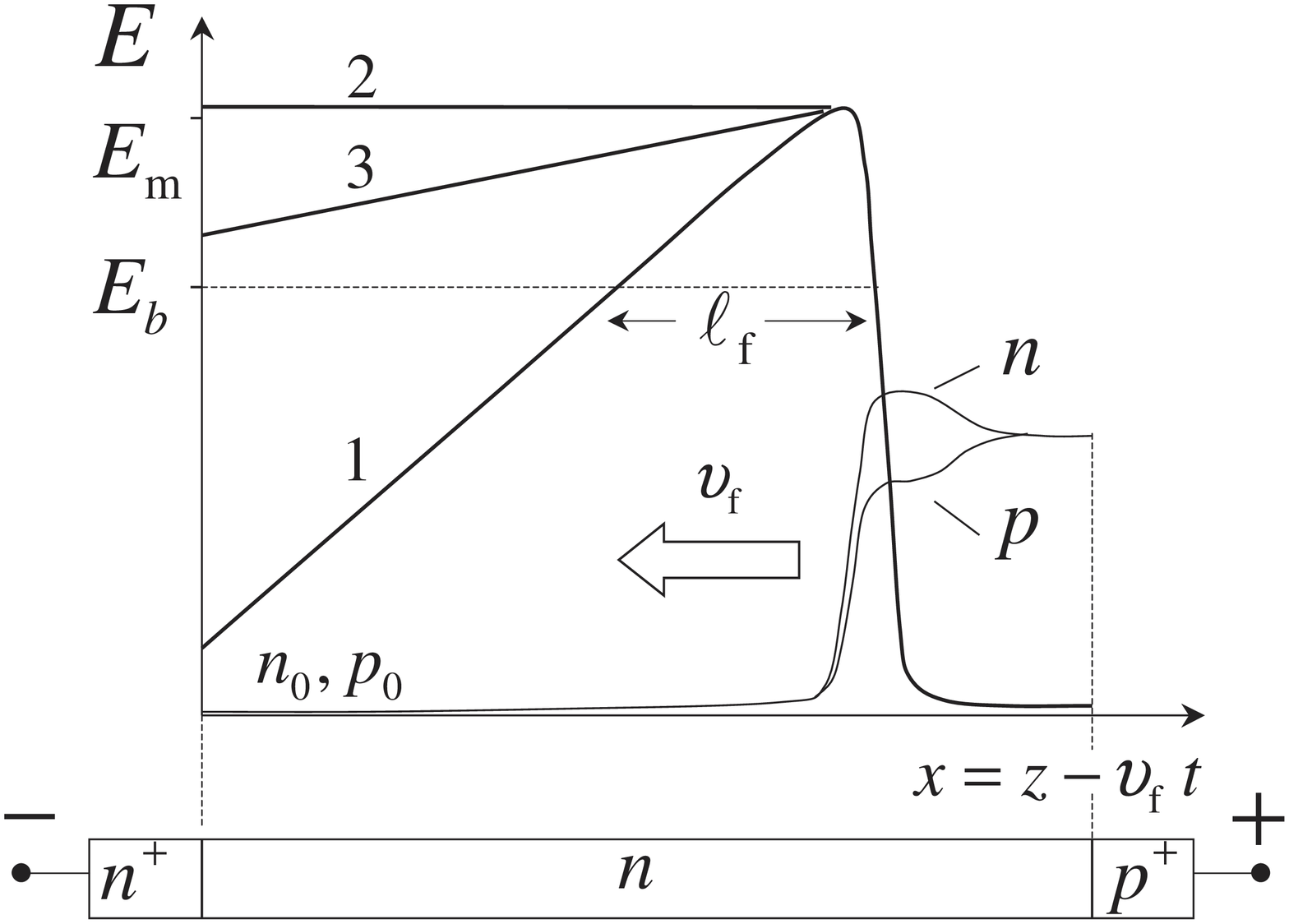}
\end{center}
\caption {Electric field profiles $E(x)$ in the traveling
ionization front. Profile 1 corresponds to the conventional
TRAPATT-like front with a finite size of impact ionization zone
$\ell_f$. Profile 2 corresponds to $p$-$i$-$n$ structure
($N_d=0$). Profile 3 corresponds to $p^+$-$n$-$n^+$ structure
with low $n$ base doping.} \label{sketch_voltage}
\end{figure}

We consider a planar impact ionization front and  describe it by
the standard drift-diffusion model
\cite{DEL70,ROD07} which consists of continuity equations for
electron and holes concentrations $n$ and $p$ and
the Poisson equation for the electric field $E$. For a self-similar
front motion with constant velocity $v_f$ these equations can be
simplified by introducing new variables $\sigma
\equiv n+p$, $\rho \equiv p-n$. \cite{ROD07} For intrinsic
semiconductor ($N_d=0$) the equations for $\sigma$, $\rho$ and
$E$ in the co-moving frame $z=x+v_f
\,t$ become
\begin{eqnarray}
\label{basic1} &&\frac{d}{d z} \left[v_f \,\sigma+v(E)\,\rho \right] - D \, \frac {d^2 \sigma}{d z^2}
= 2 \, \alpha(E) \,v(E) \, \sigma,\\
\label{basic2} &&v(E) \, \sigma + v_f \, \rho - D \frac{d \rho}{dz}=0,
\\ \label{basic3}
&&\frac{d E}{dz}=\frac{q}{\varepsilon \varepsilon_0} \, \rho,
\end{eqnarray}
where $v(E)$ is the drift velocity, $D$ is the diffusion
coefficient and $\alpha(E)$ is the impact ionization coefficient.
Here we neglect recombination and assume that
electrons and holes are identical in a sense that $\alpha(E)=\alpha_n(E)=\alpha_p(E)$
and $v(E)=v_n(E)=v_p(E)$. For an infinite domain
the boundary conditions are $E \rightarrow
E_{\rm m}$, $\sigma,\rho \rightarrow 0$ for $z \rightarrow
-\infty$ and $E, \rho \rightarrow 0$, $\sigma \rightarrow
\sigma_{\rm m}$ for $z \rightarrow +\infty$.
\begin{figure}
\begin{center}
\includegraphics[width=5.0 cm,height=6.5 cm,angle=270]{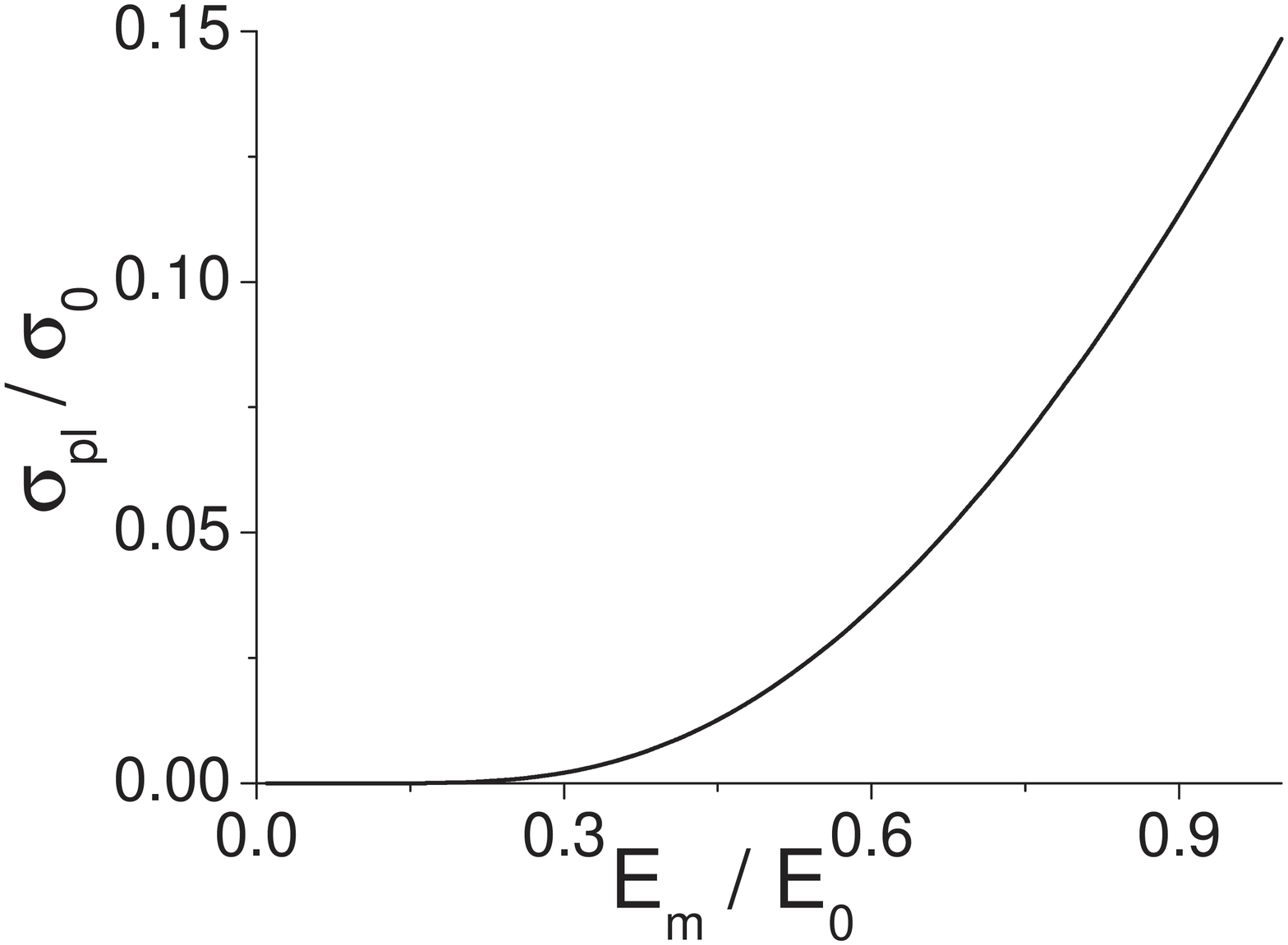}
\end{center}
\caption {Concentration of electron-hole plasma $\sigma_{\rm pl}$,
normalized by $\sigma_0 = 2 \varepsilon \varepsilon_0 \alpha_0 E_0/q$,
behind the front as a function of electric field $E_{\rm m}$
for Townsend's approximation of the impact ionization coefficient
$\alpha(E)=\alpha_0 \exp(-E_0/E)$.}
\label{concentration}
\end{figure}

In the simplified case of $D=0$ we use Eq.\ (\ref{basic2}) to exclude
$\rho$ from Eqs.(\ref{basic1}) and (\ref{basic3}). This yields
\begin{eqnarray}
\label{basic1a}
&&\frac{d}{dz}\left[\frac{v_f^2-v^2(E)}{v_f}\,\sigma \right]=2\ \alpha(E) v(E) \sigma,\\
\label{basic2a}
&&\frac{dE}{dz}=-\frac{q}{\varepsilon \varepsilon_0} \frac{v(E)}{v_f} \sigma,\\
\label{basic3a}
&&v(E) \, \sigma + v_f \, \rho=0.
\end{eqnarray}
Then by dividing equations (\ref{basic1a})
and (\ref{basic2a}) and integrating over $E$ we immediately find the dependence $\sigma(E)$
in the moving front:
\begin{equation}
\sigma(E)=\frac{2 \varepsilon \varepsilon_0}{q} \frac{v_f^2}{v_f^2-v^2(E)}
\int^{\rm E_{\rm m}}_E \alpha(E)\,dE.
\label{sigma_E}
\end{equation}
The plasma concentration far behind the ionization zone
\begin{equation}
\label{sigma_pl}
\sigma_{\rm pl} =
\frac{2 \varepsilon \varepsilon_0}{q} \int^{E_{\rm m}}_0 \alpha(E)\,dE
\end{equation}
depends only on the electric field
$E_{\rm m}$ (Fig.\ 2). The dependences $p(E)$ and $n(E)$
in the traveling front 
\begin{equation}
\label{np}
p(E),n(E)=\frac{\varepsilon \varepsilon_0}{q} \frac{v_f}{v_f \pm v(E)} \int^{E_m}_{E} \alpha(E) \, dE.
\end{equation}
follow directly from Eqs.\ (\ref{basic3a},\ref{sigma_E}) 
and are shown in Fig.\ 3.
Remarkably, a traveling front solution exists  for any $v_f \ge v_s$.
Within the approximation $D=0$ the slowest solution corresponds to the
shock front (discontinious at $E=E_{\rm m}$) that travels with a saturated
drift velocity $v_f=v_s$.

\begin{figure}
\begin{center}
\includegraphics[width=5.0 cm,height=6.5 cm,angle=270]{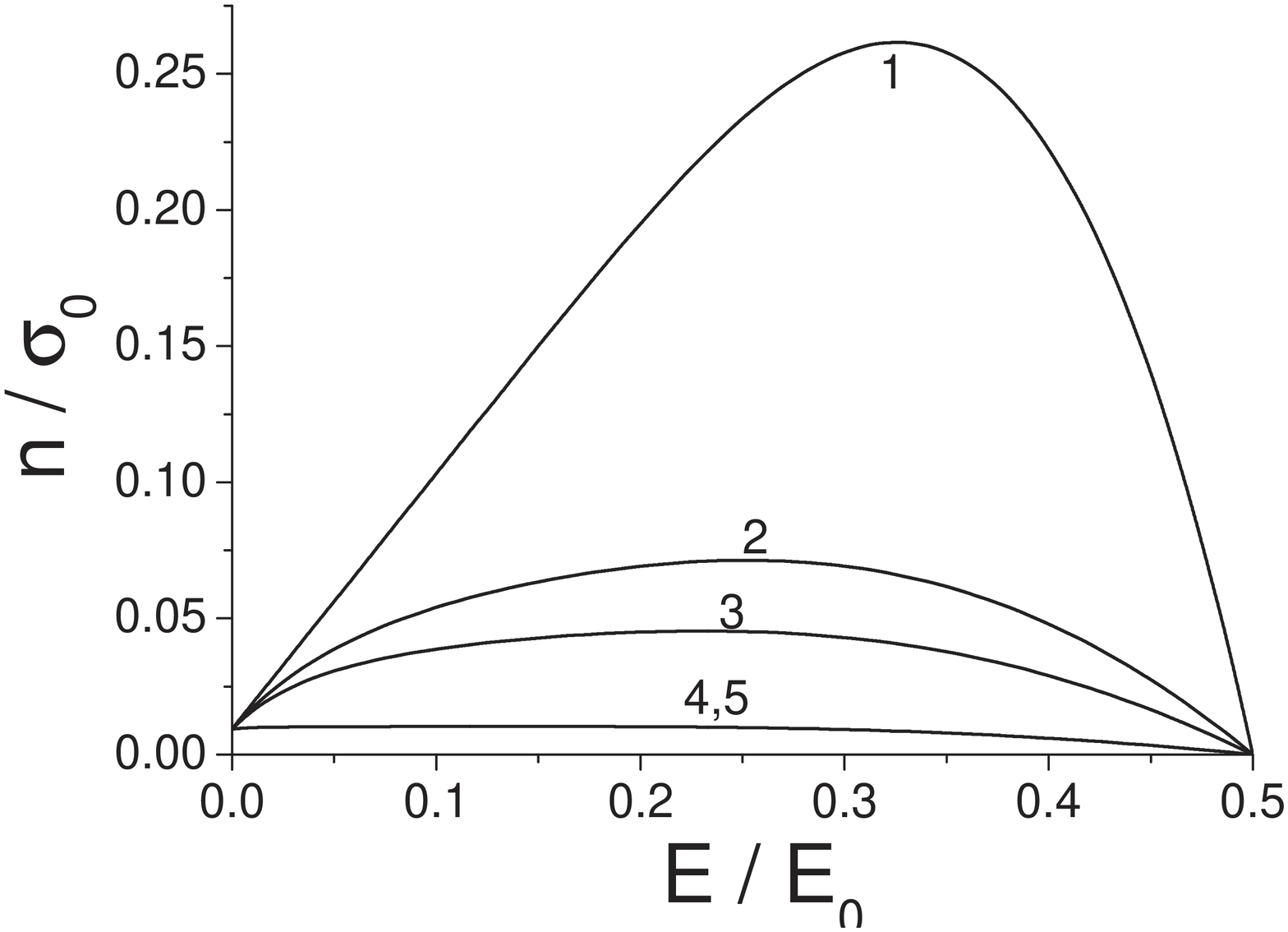}\\
\includegraphics[width=5.0 cm,height=6.5 cm,angle=270]{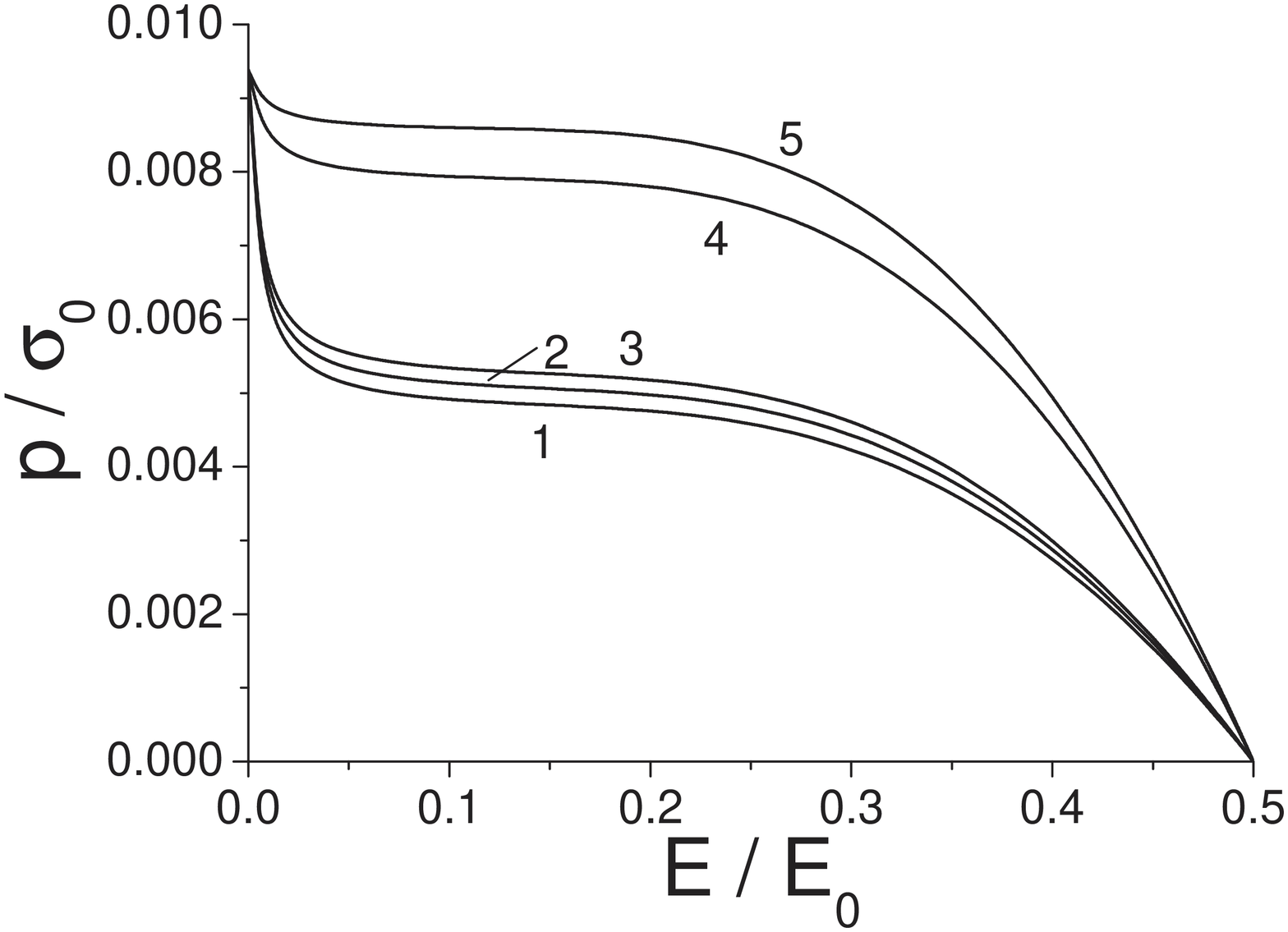}
\end{center}
\caption {Concentrations of electrons $n$ and holes $p$,
normalized by $\sigma_0 = 2 \varepsilon \varepsilon_0 \alpha_0 E_0/q$,
in the traveling front as functions of electric field $E$ for different front
velocites $v_f/v_s=1,1.1,1.2,5,10$ (curves 1,2,3,4, and 5, respectively)
and approximations $\alpha(E)=\alpha_0 \exp(-E/E_0)$,
$v(E)=v_s/(E+E_s)$; $E_s/E_0=0.01$.
The applied electric field is $E_{\rm m}=E_0/2=50 E_s$.
Note that for the chosen $v(E)$ approximation $v(E_{\rm m})=0.98 \, v_s$,
hence $v_f>v_s$ for all shown curves.}
\label{Si}
\end{figure}

The above analysis does not allow to select a physically relevant
solution and hence to find the actual front velocity $v_f$. The
selection problem remains in the case of $D \ne 0$. This is a
general feature of fronts propagating into linearly unstable state
[see Ref.~\onlinecite{SAA03} and references therein]. Ionizing
fronts belong to this class since the state $(E=E_{\rm
m},\sigma=0)$ is unstable: due to $E_{\rm m}>E_b$ any amount of
free carriers leads to avalanche multiplication. It has also been
suggested and confirmed by numerical simulations that in gases
\cite{Lag,Ute} and semiconductors \cite{KYU07} ionizing fronts are
so called {\it pulled} fronts. For a pulled front, the dynamics in
the part of the front where avalanche multiplication and screening
are essentially nonlinear is subordinated to the linear dynamics
of the front tip which fully determines the propagation
velocity.\cite{SAA03} The dynamics of the front tip is described
by the linearized  (near the state $E=E_{\rm m}$, $\sigma =0$)
version of equations (\ref{basic1},\ref{basic2}) with constant
coefficients $v(E)=v_s$ and $\alpha(E)=\alpha(E_{\rm
m})=\alpha_{\rm m}$
\begin{eqnarray}
\label{linear1a} &&v_f \, \frac{d \sigma}{d z} + v_s \frac{d \rho}{d z}
- D \frac {d^2 \sigma}{d z^2} = 2 \, v_s \,
\alpha_{\rm m} \, \sigma, \\ \label{linear1b}
&&v_s \, \sigma+v_f\,\rho - D\frac{d \rho}{dz}=0,
\; \alpha_{\rm m} \equiv \alpha(E_{\rm m}).
\end{eqnarray}
Here we take into account that in ionizing fields $v(E)=v_s={\rm const}$.
Solutions of these linear equations are exponential functions 
$\sigma(z),\rho(z)  \sim \exp (\lambda\,z)$, where the dispersion
relation $v_f(\lambda)$ remains to be found.

It is known from the theory of pulled fronts that the actual front
velocity strongly depends on the type of initial conditions.\cite{SAA03}
All possible initial conditions split in two classes that lead to qualitatively different
dynamics. {\it Localized} conditions correspond to initial profiles $\sigma(x,t=0)$
that are steeper than the profile $\exp(\lambda^{\star}x)$
with a certain characteristic exponent $\lambda^{\star}$: $\sigma(x) < C \exp(\lambda^{\star}x)$ for 
$x \rightarrow -\infty$, where $C$ is an arbitrary constant.
In this case the front profile eventually becomes smoother and
asymptotically reaches the profile $\sigma(z) \sim \exp
(\lambda^{\star} z)$ at $z \rightarrow - \infty$ that propagates
with {\it linear marginal stability velocity}
$v^{\star}=v_f(\lambda^{\star})$. \cite{SAA03}
Any initial profile that is strictly equal to zero for sufficiently
small value of $x$ also represents a localized initial condition.
{\it Nonlocalized} initial conditions correspond to
profiles $\sigma(x,t=0)$ with slow spatial decay that do not meet the above mentioned
condition $\sigma(x) < C \exp(\lambda^{\star}x)$ for 
$x \rightarrow -\infty$ and hence are smoother than $\exp(\lambda^{\star}x)$.
In this case the front velocity is fully determined by
$\sigma(x,t=0)$: for the initial profile
$\exp(\lambda_0 x)$ with $\lambda_0<\lambda^{\star}$  the front velocity is 
given by the dispersion relation $v_0=v_f(\lambda_0)$.

\begin{figure}
\begin{center}
\includegraphics[width=5.0 cm,height=6.5 cm,angle=270]{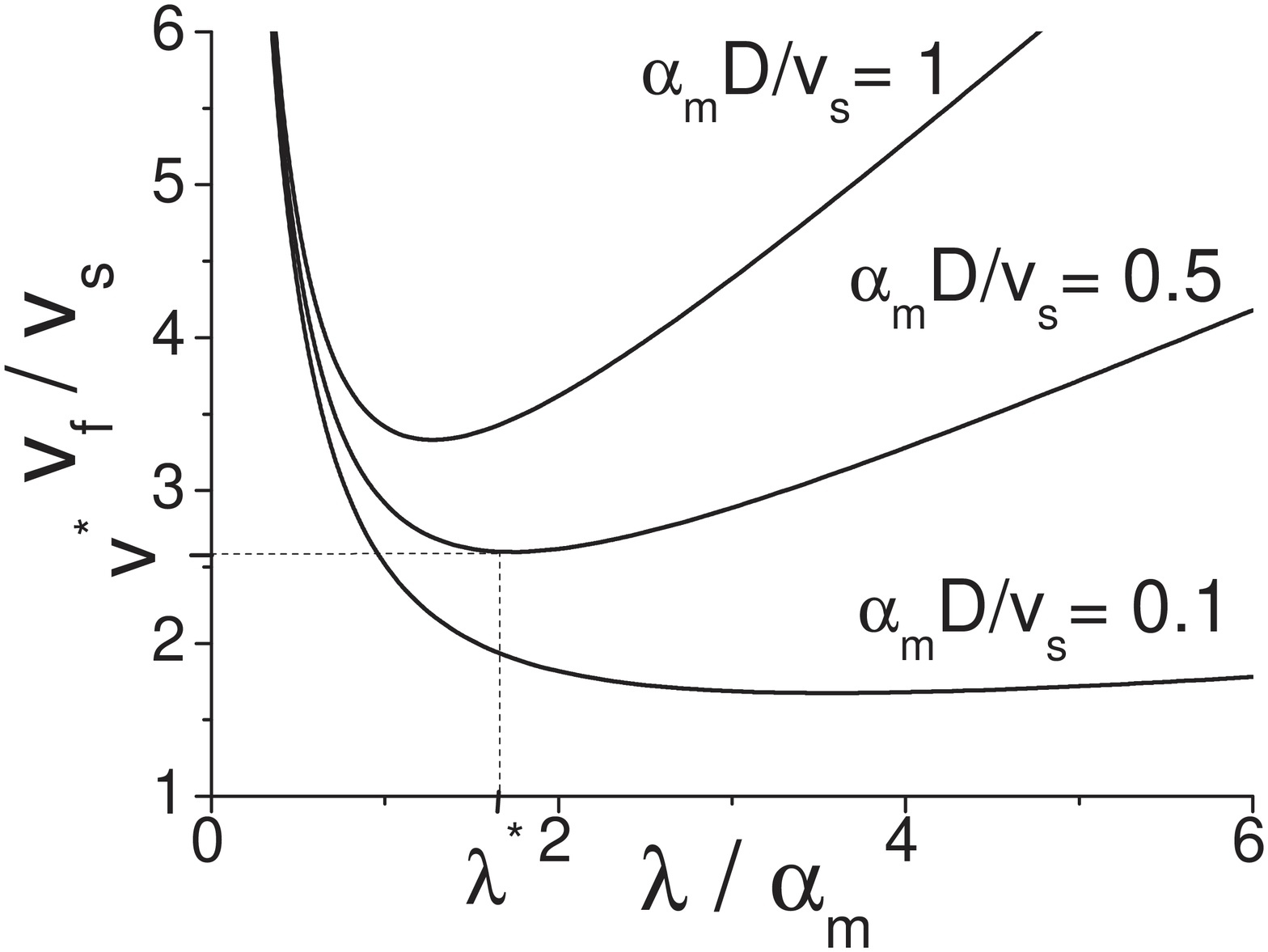}
\end{center}
\caption{The dispersion relation $v_f(\lambda)$.}
\end{figure}

The dispersion relation $v_f(\lambda)$ follows from the characterictic
equation of Eqs. (\ref{linear1a}) and (\ref{linear1b})
\begin{eqnarray}
\nonumber
\label{characteristic} \ell^2 \, \lambda^3 - 2 \ell \left(\frac{v_f}{v_s}\right)\lambda^2 
&+& \left[\left(\frac{v_f}{v_s}\right)^2-1+2 \alpha_{\rm m} \ell \right] \lambda \\
&-& 2 \alpha_{\rm m} \left(\frac{v_f}{v_s}\right) = 0, \; \ell \equiv \frac{D}{v_s}
\end{eqnarray}
and is explicitly given by (see Fig.\ 4)
\begin{equation}
\label{vfk} \frac{v_f(\lambda)}{v_s}= \frac{\alpha_{\rm m}}{\lambda} +
\sqrt{1+\left(\frac{\alpha_{\rm m}}{\lambda}\right)^2}+
\ell \,\lambda.
\end{equation}
\\
The critical steepness $\lambda^{\star}$ and the velocity $v^{\star}$
correspond to the minimum point \cite{SAA03} and are given by
\\
\begin{eqnarray}
\frac{\lambda^{\star}}{\alpha_{\rm m}}&=&
\sqrt{\frac{1}{\alpha_{\rm m} \ell} \left(1-\frac{\alpha_{\rm m}\ell}{2}+A\right) },\\
\frac{v^{\star}}{v_s}&=&
\sqrt{1+5\alpha_{\rm m}\ell-\frac{(\alpha_{\rm m} \ell)^2}{2}+
(4+\alpha_{\rm m}\ell) A}, \\
\nonumber
&&{\rm where} \qquad A \equiv \sqrt{(\alpha_{\rm m} \ell)^2/4+\alpha_{\rm m}\ell}.
\end{eqnarray}

The right branch of the $v_f(\lambda)$
($\lambda>\lambda^{\star}$) corresponds to fronts whose velocity
increases with steepness $\lambda$ due to diffusion. According to
the concept of localized initial conditions,\cite{SAA03} these
fronts are unstable: their steep profiles eventually relax to the
profile with exponential tip $\exp[\lambda^{\star}\,z]$ that
propagates with the velocity $v^{\star}$. The left branch of the
$v_f(\lambda)$ dependence ($\lambda<\lambda^{\star}$) corresponds
to stable fronts whose velocity decreases with $\lambda$. These
fronts correspond to nonlocalized initial conditions which are in
the focus of our interest.

\begin{figure}
\begin{center}
\includegraphics[width=5.0 cm,height=6.5 cm,angle=270]{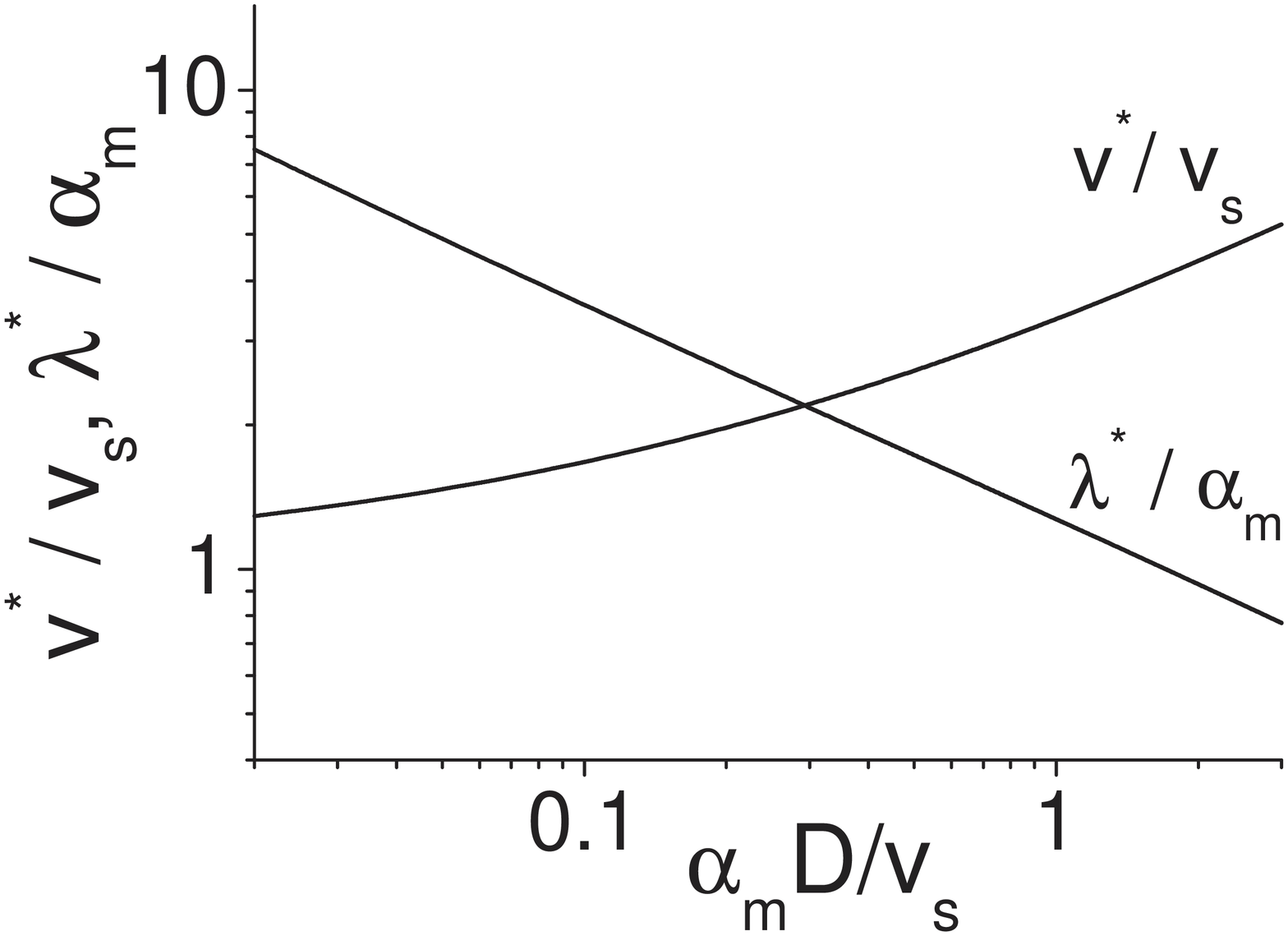}
\end{center}
\caption{Linear marginal velocity $v^{\star}$ and the steepness $\lambda^{*}$.}
\end{figure}

Charateristic values  of the dimensionless parameter 
$\alpha_{\rm m} \ell \equiv \alpha_{\rm m} D/v_s$ 
are 0.1 for Si
and 1 for GaAs devices. As follows from Fig.\ 5, $\lambda^{\star} >
\alpha_{\rm m}$ in the relevant interval. Physically it  means that the
front propagating with linear marginal stability velocity
$v^{\star}$ is so steep that the validity of the drift-diffusion
approximation is questionable. This problem disappears for much
smoother fronts that correspond to nonlocalized initial conditions
(left branch in Fig.\ 4) if $\lambda<\lambda^{\star}$.

Eq.\ (\ref{vfk}) leads to a simple
relation ${v_f}/{v_s}={2 \alpha_{\rm m}}/{\lambda}$
for the ionizing front velocity in case of preionization
with decay exponent $\lambda < \lambda^{\star}$.
For such fast fronts the effect of diffusion is negligible; in particular,
Eq.\ (\ref{basic1a},\ref{basic2a},\ref{basic3a},\ref{sigma_E},\ref{sigma_pl},\ref{np})
are fully applicable.
Although $v_f$ increase with $\alpha_{\rm m}$ and hence with
electric field $E_{\rm m}$, it is the ratio $\alpha_{\rm m}/\lambda$
which actually counts. This ratio can be arbitrarily large resulting in front
velocities that exceed the saturated drift velocity by many
orders of magnitude.  It means that a proper choice of slowly
decaying preionization profile gives the possibility to achieve
fast front propagation in even moderate (with respect to $E_b$) electric fields.
However, the concentration of electron-hole plasma generated by
the front passage increases with $E_{\rm m}$ (Fig.\ 2).
Generally, the electromagnetic limitation $v_f < c$, where $c$
is the velocity of light, may be important: for $v_f$ comparable to
$c$ the full set of Maxwell equations shall substitute the
Poisson equation in the model because
the feedback from a nonstationary electromagnetic field
created by the front passage on the front dynamics becomes essential.

Pre-ionization profiles with slow spatial decay can appear due to
photoionization by photons from dense electron-hole plasma
behind the front.
In this case, $\lambda^{-1}$ can be roughly identified as the light absorption
length. This mechanism is efficient
in direct-band materials and can be relevant for
planar fronts in GaAs diode structures \cite{GaAs} as well as
finger-like streamers in direct-band bulk semiconductors.\cite{DYA}
Another mechanism is related to field-enhanced ionization
of deep centers in Si $p^{+}$-$n$-$n^{+}$  structures
used in pulse power applications. \cite{Si}
These high-voltage structures
possess ``hidden'' deep levels -- process-induced defects --
with low recombination activity. \cite{sulfur}
Preionization of the high-field space charge region can be due to
field-enhanced ionization of these deep centers embedded in the $n$
base.\cite{ROD05}  This ionization
is more efficient near the $p^{+}$-$n$ junction where the electric field
is stronger. Hence the profile of initial carriers
decreases along the $n$ base. The characteristic decay length
$\lambda^{-1}$ is expected to be a fraction of the $n$ base width $W \sim 100 \;{\rm
\mu m}$. At the same time for low doped $n$ base
the electric field can be above the ionization threshold $E_b$ everywhere.
This may result in front velocities that exceed $v_s$ by
several orders of magnitude. Numerical simulations of such triggering
process will be presented elsewhere.

Acknowledgements.-- We are indebted to U.~Ebert and
V.~Kachorovskii for  enlightening discussions. This work was supported
by the Programm of Russian Academy of Sciences, ``Power semiconductor
electronics and pulse technologies".
P.~Rodin is grateful to A. Alekseev for hospitality at the University
of Geneva and acknowledges the support from the Swiss National
Science Foundation.

\end{document}